\documentclass[11pt,superscriptaddress,aps,prd,preprint]{revtex4}
\usepackage{amsmath}

\makeatletter
\usepackage[T1]{fontenc}
\usepackage{amsmath}
\usepackage{amssymb}
\usepackage{graphicx}

\usepackage{slashed}

\DeclareSymbolFont{extraitalic}      {U}{zavm}{m}{it}
\DeclareMathSymbol{\Qoppa}{\mathord}{extraitalic}{161}
\DeclareMathSymbol{\qoppa}{\mathord}{extraitalic}{162}
\DeclareMathSymbol{\Stigma}{\mathord}{extraitalic}{167}
\DeclareMathSymbol{\Sampi}{\mathord}{extraitalic}{165}
\DeclareMathSymbol{\sampi}{\mathord}{extraitalic}{166}
\DeclareMathSymbol{\stigma}{\mathord}{extraitalic}{168}

\newcommand{\bea}{\begin{eqnarray}}
\newcommand{\eea}{\end{eqnarray}}

\begin{document}

\title{G\"{o}del and G\"{o}del-type universes in K-essence theory}

\author{J. G. da Silva}\email[]{junior@fisica.ufmt.br }
\affiliation{Instituto de F\'{\i}sica, Universidade Federal de Mato Grosso,\\
78060-900, Cuiab\'{a}, Mato Grosso, Brazil}

\author{A. F. Santos}\email[]{alesandroferreira@fisica.ufmt.br}
\affiliation{Instituto de F\'{\i}sica, Universidade Federal de Mato Grosso,\\
78060-900, Cuiab\'{a}, Mato Grosso, Brazil}

\begin{abstract}

In this paper, the G\"{o}del-type solutions within the k-essence theory are investigated. The consistency of field equations, causality violation and existence of closed timelike curves are studied.  The conditions for existence of G\"{o}del and G\"{o}del-type solutions are determined for any function that defines the k-essence. By taking different sources of matter, it is shown that this theory of gravity allows causal and non-causal G\"{o}del-type solutions.

\end{abstract}

\maketitle

\section{Introduction}

Observational data of Type Ia supernovae \cite{Riess, Per}, cosmic microwave background \cite{Spergel} and large scale structure \cite{Teg} suggest that the universe is undergoing cosmic acceleration. The accelerated expansion of the universe can be understood as being due to an exotic energy density with negative pressure, responsible by two-third of the universe total energy. This exotic component is called dark energy. Dark energy has become one of the most important problems of the modern cosmology. Although it is intensely studied, little is known about dark energy. One candidate for such a component is a cosmological constant ($\Lambda$). This candidate is in agreement with the current observations, but presents difficulties in reconciling the small observational value of dark energy density with estimates from quantum field theory. This is a well-known problem, the cosmological constant problem \cite{Weinberg, Carroll}. Other candidates for dark energy are the quintessence \cite{quint}, Chaplygin gas \cite{Pas}, tachyon \cite{Sen00} and k-essence \cite{Picon1}. Here the k-essence is considered. 

K-essence is an approach to explain the accelerated expansion of the universe. The main idea of k-essence has been motivated from the Born-Infeld action of string theory \cite{Sen0, Lam, Callan}. The first ideas were used to introduce a possible model for explaining inflation \cite{Arm, Gar, Picon}. After some studies, it was noted that the k-essence could also yield interesting models for the dark energy \cite{Picon1, Picon2, Chimento, Chimento1}. It works with the idea that the unknown dark energy component is due exclusively to a minimally coupled scalar field with non-canonical kinetic energy which results in the negative pressure. There is a subset of k-essence that contains purely kinetic k-essence, i.e., the Lagrangian contains only a kinetic term and does not depend explicitly on the scalar field itself. These models have been investigated in the context of inflation.  The k-essence has been studied in several different contexts, such as: the FRW k-essence cosmologies has been analyzed \cite{Agui}, exact solutions in K-essence for isotropic cosmology have been studied \cite{Pimentel}, the stability of the model has been investigated \cite{Nelson}, the dynamics of k-essence has been discussed \cite{AD}, the slow-roll conditions for k-essence has been obtained \cite{Chiba}, a connection between the holographic dark energy density and the kinetic k-essence energy density has been studied \cite{Cruz}, dark matter to dark energy transition in k-essence cosmologies has been examined \cite{Forte}, thermodynamic properties of k-essence has been discussed \cite{bilic}, power-law expansion in k-essence cosmology has been investigated \cite{Feins}, an attempt to unify dark matter and dark energy using kinetic k-essence has been done \cite{Sherrer}, holographic dark energy in Bianchi-III universe with k-essence has been analyzed \cite{Bianchi},  k-essence in the framework of Horndeski gravity has been studied \cite{Horn}, quantum cosmology with k-essence has been discussed \cite{Fabris}, among others. In this paper the consistence of the G\"{o}del and G\"{o}del-type universes within k-essence is investigated, as well as its consequences.

The G\"{o}del metric \cite{Godel} is a homogeneous rotating cosmological solution of General Relativity (GR)  with pressureless matter and negative cosmological constant, which played an important role in the conceptual development of GR. This solution leads to the possibility of Closed Timelike Curves (CTCs), which allow violation of causality and makes time-travel theoretically possible in this space-time. CTCs are also admitted in other cosmological models, such as in the Van-Stockum model, cosmic string, Kerr black hole \cite{Van, Cosmic, Kerr}, among others. A generalization of the G\"{o}del solution (G\"{o}del-type solution) has been developed \cite{Reboucas}. In the G\"{o}del-type universes both causal and non-causal regions can be explored. These regions are determined from free parameters of the metric. Furthermore, the problem of causality may be examined with more details and three different classes of solutions are defined. These solutions are classified as: (i)  causal solution, where the breakdown of causality of G\"{o}del-type is avoided, (ii) regions with an infinite sequence of alternating causal and non-causal solutions, and (iii) there is only one non-causal region. The compatibility of G\"{o}del and G\"{o}del-type solutions and its consequences have been investigated in several modified theories of gravity. Some examples are, Chern-Simons gravity \cite{CS, CS1}, $f(R)$ gravity \cite{fR, fR1}, $f(T)$ gravity \cite{fT}, $f(R,T)$ gravity \cite{fRT, fRT1}, Horava-Lifshitz gravity \cite{Hor}, bumblebee gravity \cite{Bum}, Brans-Dicke theory \cite{BD} and $f(R,Q)$ gravity \cite{fRQ}.

This paper is organized as follows. In section II, a brief introduction to k-essence theory is presented. In section III, the G\"{o}del and G\"{o}del-type universes are discussed. In section IV, the G\"{o}del metric in k-essence theory is analyzed. In section V, the G\"{o}del-type universe in k-essence for different matter contents is studied. For the perfect fluid as matter source, non-causal G\"{o}del-type solution is found. Causal solutions emerge for a combined matter source of perfect fluid and electromagnetic field or for a single electromagnetic field. In section VI, some concluding remarks are discussed.

\section{K-essence theory}

In this section a brief introduction to the k-essence theory is presented.  The k-essence theory arose in string theory \cite{Callan}. In cosmology, it was studied for the first time in the context of inflation \cite{Picon} and then it was suggested as dynamical dark energy \cite{Picon1}. There is also an attempt to describe dark matter using k-essence \cite{Sen}. A possibility for achieving a triple unification, i.e., inflation, dark matter and dark energy within the context of the same model, has been explored \cite{Bose}. The k-essence belong a class of models that contains non-canonical kinetic  terms in the Lagrangian. The action that describes the k-essence theory is given by
\begin{equation}
    S=\int d^4x \sqrt{-g} \ \left(\frac{R}{2\beta^2} - K(\phi, X) + \mathcal{L}_m \right),
\end{equation}
where $g$ is the metric determinant, $R$ is the Ricci scalar, $\beta^2=8\pi G$, $\mathcal{L}_m$ is the matter Lagrangian and $K(\phi, X)$ is a function that depends on scalar field $\phi$ and of its derivatives, since $X\equiv\partial^\mu\phi\partial_\mu\phi$.

Varying the action with respect to the metric $g_{\mu\nu}$, the field equation is given as
\begin{equation}
    \frac{1}{\beta^2} \left(R_{\mu\nu} - \frac{1}{2} \ g_{\mu\nu} \ R \right) = K(\phi,X) g_{\mu\nu} - 2K_x(\phi,X) \ \partial_\mu \phi \partial_\nu \phi + T_{\mu\nu},\label{fe1}
\end{equation}
where $K_x(\phi,X) = \frac{\partial K(\phi,X)}{\partial X}$ and $T_{\mu\nu}$ is the energy-momentum tensor of the matter content that is defined as
\begin{equation}
 T_{\mu\nu} = -\frac{2}{\sqrt{-g}}  \frac{\delta \left(\sqrt{-g}{\cal L}_m\right)}{\delta g^{\mu\nu}}. \label{TEM}
\end{equation}
By varying the action with respect to the scalar field gives the second field equation
\begin{equation}
    - K_{\phi}(\phi,X) + 2\nabla_{\mu} \left(K_{x}(\phi,X) \partial^{\mu} \phi \right)=0,
\end{equation}
with $K_\phi(\phi,X) = \frac{\partial K(\phi,X)}{\partial \phi}$ and $\nabla^{\mu}$ being the covariant derivative. Using the covariant derivative definition in terms of connection coefficients, the field equation becomes
\begin{equation}
    - K_{\phi}(\phi,X) + 2\partial_{\mu} \left(K_{x}(\phi,X) \partial^{\mu} \phi \right) + 2 \Gamma^{\mu}_{\mu\rho} \left(K_{x}(\phi,X) \partial^{\rho} \phi \right) = 0,
\end{equation}
where $\Gamma^{\mu}_{\mu\rho}$ is the connection coefficient.

In the next section the main objective is to investigate whether  the G\"{o}del-type solutions is consistent with the k-essence theory.

\section{G\"{o}del and G\"{o}del-type universes}

Kurt F. G\"{o}del presented, in 1949,  an exact solution of the Einstein field equations, called the G\"{o}del universe \cite{Godel}. It is the first cosmological solution with rotating matter. In addition,  it is spatially homogeneous, stationary, possessing cylindrical symmetry, and exhibit the possibility of the Closed Timelike Curves (CTCs). As a
consequence, a traveler moving along such curves can come back to his own past leading to causality violation.

The line element that describes the G\"{o}del universe \cite{Godel} is
\begin{equation}
    ds^2 = a^2 \left(dt^2 - dx^2 + \frac{e^{2x}}{2}dy^2 - dz^2 + 2e^xdt \ dy \right), 
\end{equation}
where $a$ is an arbitrary number. The metric components are given explicitly as
\begin{equation}
    g_{\mu\nu} = \left(\begin{array}{cccc}
    1   & 0    & e^x  & 0 \\
    0  & -1 & 0 & 0  \\
    e^x   & 0 & \frac{1}{2}e^{2x} & 0 \\
    0 & 0 & 0 & -1\end{array}\right).
\end{equation}
The non-zero Ricci tensor components are
\bea
R_{00} = 1 ;\quad  R_{02} = e^x ; \quad R_{22} = e^{2x},  
\eea
and the Ricci scalar is 
\bea
R = \frac{1}{a^2}.
\eea

To solve Einstein equations G\"{o}del considered that the energy-momentum tensor is composed of two terms: a matter density $\rho$, that represent a homogeneous distribution of dust particles and a non-zero cosmological constant $\Lambda$ . Then he found that the field equations should satisfy the conditions \cite{Godel}
\bea
\Lambda=-\frac{1}{2a^2}\quad \mathrm{and}\quad 8\pi G\rho=\frac{1}{a^2}.
\eea
This solution leads to the G\"{o}del metric to be an exact solution of the Einstein theory of gravity.  So this implies that general  relativity allows causality violation.

In order to obtain more information on the issue of causality, a generalization of the G\"{o}del solution,  called G\"{o}del-type metrics, has been proposed \cite{Reboucas}. Its line element is given as
\begin{equation}
    ds^2= \left[dt + H(r) d\phi   \right]^2 - D^2(r) d\phi^2 - dr^2 -dz^2,\label{godel}
\end{equation}
where the functions $H(r)$ and $D(r)$ must obey the relations
\bea
\frac{H'(r)}{D(r)}&=&2\omega,\label{7}\\
\frac{D''(r)}{D(r)}&=&m^2.\label{8}
\eea
The prime denotes the derivative with respect to $r$, and $\omega$ and $m$ are free parameters. The solutions of eqs. (\ref{7}) and (\ref{8}) may be divided into three different classes of G\"{o}del-type metrics in terms of $m^2$: (i) hyperbolic class ($m^2>0$), (ii) trigonometric class ($m^2<0$) and (iii) linear class ($m^2=0$). Here the hyperbolic class is considered. In this class the functions $H(r)$ and $D(r)$ are defined as
\bea
    H(r) &=& \frac{4\omega}{m^2} senh^2 \left(\frac{mr}{2} \right) \label{9}\\
    D(r) &=&\frac{1}{m} senh(mr).\label{10}
\eea

A particular case of the hyperbolic class is $m^2=2\omega^2$, which correspond to the G\"{o}del solution. An important characteristic of the G\"{o}del-type solution is the possibility of existence of CTCs, that are circles defined by $t, z, r = \rm{const}$. The causality violation occur for $r > r_c$ such that
\begin{equation}
 \sinh^2\left(\frac{mr_c}{2}\right) = \left(\frac{4\Omega^2}{m^2}-1\right)^{-1},\label{CR}
\end{equation}
where $r_c$ is the critical radius. Note that $m^2 =  4\Omega^2$ leads to $r_c\rightarrow\infty$. As a consequence, causality is not violated.

Using the functions defined in eqs. (\ref{9}) and (\ref{10}), the metric components are written, explicitly, as
\begin{equation}
     g_{\mu\nu} =\left(\begin{array}{cccc} 
        1 & 0 & \frac{4\,w\,{\mathrm{sinh}^2\left(\frac{m\,r}{2}\right)}}{m^2} & 0\\ 0 & -1 & 0 & 0\\ \frac{4\,w\,{\mathrm{sinh}^2\left(\frac{m\,r}{2}\right)}}{m^2} & 0 & \frac{16\,w^2\,{\mathrm{sinh}^4\left(\frac{m\,r}{2}\right)}}{m^4}-\frac{{\mathrm{sinh}^2\left(m\,r\right)}}{m^2} & 0\\ 0 & 0 & 0 & -1 
    \end{array}\right).
\end{equation}
Then the non-zero Einstein tensor components are
{\small
\bea
   G_{\mu\nu}= \left(\begin{array}{cccc}
       3\,w^2-m^2 & 0 & -\frac{4\,w\,{\mathrm{sinh}\left(\frac{m\,r}{2}\right)}^2\,\left(m^2-3\,w^2\right)}{m^2} & 0\\ 0 & w^2 & 0 & 0\\ -\frac{4\,w\,{\mathrm{sinh}\left(\frac{m\,r}{2}\right)}^2\,\left(m^2-3\,w^2\right)}{m^2} & 0 & \frac{4\,w^2\,{\mathrm{sinh}\left(\frac{m\,r}{2}\right)}^2\,\left(-3\,m^2\,{\mathrm{sinh}\left(\frac{m\,r}{2}\right)}^2+m^2+12\,w^2\,{\mathrm{sinh}\left(\frac{m\,r}{2}\right)}^2\right)}{m^4} & 0\\ 0 & 0 & 0 & 
       m^2-w^2 
    \end{array}\right),\label{13}
\eea}
and the Ricci scalar is
\begin{equation}
    R = 2(m^2 - \omega^2).
\end{equation}

For simplicity, let's choose a new basis such that the metric becomes
\begin{equation}
    ds^2 = \eta_{AB} \theta^A \theta^B = (\theta^0)^2 - (\theta^1)^2 - (\theta^2)^2 - (\theta^3)^2, \label{frame}
\end{equation}
where
\bea
    \theta^{(0)} &=& dt + H(r)d\phi \\ \theta^{(1)} &=& dr \\ \theta^{(2)} &=& D(r)d\phi \\ \theta^{(3)} &=& dz,
\eea
with $\theta^A \equiv e^A\ _{\mu} dx^\mu$ and $e^A\ _{\mu}$ are the tetrads, which the non-null components are 
\begin{equation}
    e^{(0)}\ _{0} = e^{(1)}\ _{1} = e^{(2)}\ _{2} = 1 , \quad e^{(0)}\ _{2} = H(r), \quad e^{(2)}\ _{2} = D(r).
\end{equation}
Note that, the latin indices represent the flat space-time and they are used between brackets. Using that $e^\mu\,_A$ is defined by $e^A\,_\mu e^\mu\,_B=\delta^A_B$, we get
\begin{equation}
        e^{0}\ _{(0)} = e^{1}\ _{(1)} = e^{3}\ _{(3)} = 1, \quad e^{0}\ _{(2)} = - \frac{H(r)}{D(r)}, \quad e^{2}\ _{(2)} = D^{-1}(r).
\end{equation}

Then the non-vanishing components of the Einstein tensor (\ref{13}) in the flat (local) space-time take the form
\bea
    G_{(0)(0)} &=& 3\omega^2-m^2 \\
    G_{(1)(1)} &=& G_{(2)(2)} = \omega^2 \\
    G_{(3)(3)} &=& m^2-\omega^2,
\eea
where $G_{AB}=e^\mu_A e^\nu_B G_{\mu\nu}$ has been used.

\section{G\"{o}del universe in k-essence theory} 

Here the matter content is represented by the energy-momentum tensor defined as
\bea
T_{\mu\nu}=\rho u_\mu u_\nu + \Lambda g_{\mu\nu},
\eea
with $\rho$ being the energy density of the fluid of matter, $\Lambda$ being the cosmological constant and $u$ being a unit time-like vector whose explicit contravariant components look like $u^\mu=(\frac{1}{a}, 0, 0,0)$ and the corresponding covariant components are $u_\mu=(a, 0, ae^x, 0)$.

Then considering the G\"{o}del metric, eq. (\ref{godel}), the field equations (\ref{fe1}) become
\bea
    \beta^2 \left( K(\phi,X) \ a^2-K_x \dot{\phi}^2+a^2\rho \right) &=& \frac{1}{2}-\Lambda a^2\\
    -\beta^2 K(\phi,X)a^2  &=& \frac{1}{2} + \Lambda a^2\\
    \beta^2 \left(K(\phi,X)a^2 + a^2\rho \right)  &=& \frac{1}{2} - \Lambda a^2\\
    \beta^2 \left(\frac{1}{2}K(\phi,X)a^2 + a^2\rho \right)  &=& \frac{3}{4} - \frac{1}{2} \Lambda a^2.
\eea

This set of equations is satisfied if
\begin{equation}
    \rho = \frac{1}{8\pi G a^2},  
\end{equation}
and
\begin{equation}
    \Lambda = -\frac{1}{2a^2} - 8\pi G \  K(\phi,X).
\end{equation}
This result implies that the G\"{o}del metric solves the modified Einstein equations if and only if these conditions are satisfied. Therefore, the k-essence theory allows causality violation for any function $K(\phi,X)$. In addition, when $K(\phi,X) \rightarrow 0 $ the  result obtained by G\"{o}del in general relativity is recovered.
 
In the next section a generalization of the G\"{o}del solution is considered. In this solution the causality issue can be investigated with more details.  Furthermore, there is a possibility to find causal and non-causal G\"{o}del-type solutions in this gravitational model.

\section{G\"{o}del-type solution in k-essence theory}

In this section the field equations of the k-essence theory are analyzed in the framework of the G\"{o}del-type metric. In order to make calculations simpler, the equations are written in the tangent (flat) space-time, i.e.,
\begin{equation}
    G_{AB}  =\beta^2\left[ K(\phi,X) g_{AB} - 2K_x(\phi,X) \ \partial_A \phi \partial_B \phi + T_{AB}\right].
\end{equation}

Let's consider the perfect fluid as the matter content such that its energy-momentum tensor is 
\bea
T_{AB}=(\rho+p)u_A u_B+pg_{AB}+\Lambda g_{AB},
\eea
where $\rho$ is the energy density of the fluid, $p$ is the pressure, $u_A$ is the 4-velocity of the matter and $\Lambda$ is the cosmological constant. As the simplest example, we assume the scalar field to be only time dependent, i.e., $\phi=\phi(t)$. Then the components of the field equations become
\bea
    \beta^2 (K(\phi,X)-2K_x(\phi,X) \dot{\phi^2} + \rho - \Lambda ) &=& 3\omega^2-m^2,\label{a1}\\
    \beta^2(K(\phi,X)-p+\Lambda) &=& \omega^2,\label{a2}\\
    \beta^2(K(\phi,X)-p+\Lambda) &=& m^2-\omega^2.\label{a3}
\eea
Equations (\ref{a2}) and (\ref{a3}) give us
\bea
    2\omega^2-m^2 = 0.
\eea
Therefore this equation leads to $m^2=2\omega^2$, which defines the G\"{o}del solution. Furthermore, these equations provide
\bea
    p&=&K(\phi,X)+\Lambda - \frac{m^2}{2\beta^2},\label{a4}\\
    \rho &=& -K(\phi,X)+\Lambda+\frac{m^2}{2\beta^2}+2K_x(\phi,X)\dot{\phi}^2.\label{a5}
\eea
Then this solution allows causality violation in this gravitational model. In other words, non-causal G\"{o}del circles (i.e., CTCs) are permitted. Thus a critical radius, which determines the existence of causal and non-causal regions, is defined as
\bea
r_c=\frac{2}{m}\sinh^{-1}(1).
\eea
Using the eqs. (\ref{a4}) and (\ref{a5}) we have
\bea
m = \beta \sqrt{\rho - p + 2K(\phi,X) -2K_x(\phi,X) \dot{\phi}^2}.
\eea
Then the critical radius becomes
\bea
r_c =  \frac{2\sinh^{-1}(1)}{\beta \sqrt{\rho - p + 2K(\phi,X) -2K_x(\phi,X) \dot{\phi}^2}}.
\eea
Therefore beyond it exists non-causal G\"{o}del circles. It depends on the k-essence theory and the matter content. Note that, the expression for the critical radius holds for any function $K(\phi,X)$.

By taking the perfect fluid as matter source a non-causal G\"{o}del-type solution is inevitably. Now a question arises: could another source of matter generate a causal G\"{o}del-type solution? Here, let's consider as matter content: (i) a combination of a perfect fluid and an electromagnetic field and (ii) a single electromagnetic field. 

\subsection{Perfect fluid plus electromagnetic field}

The matter content is composed of a perfect fluid plus an electromagnetic field aligned on z-axis and dependent of z. For this choice, the non-vanishing components of the electromagnetic tensor in frame (\ref{frame}) are
\bea
F_{(0)(3)}=E(z)\quad\mathrm{and}\quad F_{(1)(2)}=B(z),
\eea
where $E(z)$ and $B(z)$ are solutions of the Maxwell equations given by
\bea
E(z)&=&E_0\cos\left[2\omega(z-z_0)\right],\\
B(z)&=&E_0\sin\left[2\omega(z-z_0)\right],
\eea
where $E_0$ is the amplitude of the electric and magnetic fields and $z_0$ is an arbitrary constant. Thus, the non-vanishing components of energy-momentum tensor for this electromagnetic field are
\bea
T_{(0)(0)}^{EM} = T_{(1)(1)}^{EM} = T_{(2)(2)}^{EM} = \frac{E_0^2}{2}, \quad\quad T_{(3)(3)}^{EM} = - \frac{E_0^2}{2}. 
\eea
As a consequence, the total energy-momentum tensor becomes
\bea
T_{AB}=(\rho+p)u_A u_B+pg_{AB}+\Lambda g_{AB}+T_{AB}^{EM}.
\eea

Then the non-zero components of the field equations for this combined matter source are
\bea
\beta^2 (K(\phi,X)-2K_x(\phi,X) \dot{\phi^2} + \rho - \Lambda + \frac{E_0^2}{2} ) &=& 3\omega^2-m^2,\\
\beta^2(K(\phi,X)-p+\Lambda + \frac{E_0^2}{2}) &=& \omega^2,\\
\beta^2(K(\phi,X)-p+\Lambda - \frac{E_0^2}{2}) &=& m^2-\omega^2.
\eea
These field equations imply the relations
\bea
\beta^2 E_0^2 &=& 2\omega^2-m^2, \label{52}\\
 p &=& K(\phi,X)+\Lambda - \frac{m^2}{2\beta^2}.\\
 \rho &=& -K(\phi,X) + 2K_x(\phi,X) \dot{\phi^2} + \Lambda - \frac{m^2}{2\beta^2} - E_0^2. 
\eea
By assuming $E_0^2<0$, eq. (\ref{52}) permits a causal G\"{o}del-type solution with $m^2=4\omega^2$. This condition leads to $r_c\rightarrow \infty$. Therefore this combination of matter source implies no violation of causality for any k-essence function $K(\phi,X)$ that satisfies the condition $E_0^2<0$. 

\subsection{Electromagnetic field}

Here a single electromagnetic field aligned on z-axis and dependent of z is considered as matter source. The field equations become
\bea
\beta^2 (K(\phi,X)-2K_x(\phi,X) \dot{\phi^2} + \frac{E_0^2}{2} ) &=& 3\omega^2-m^2,\label{55}\\
\beta^2(K(\phi,X) + \frac{E_0^2}{2}) &=& \omega^2,\label{56}\\
\beta^2(K(\phi,X)- \frac{E_0^2}{2}) &=& m^2-\omega^2.\label{57}
\eea
Equations (\ref{56}) and (\ref{57}) lead to
\bea
\beta^2E_0^2 &=& 2\omega^2-m^2.
\eea
This is the same condition obtained from the combined matter source between perfect fluid and electromagnetic field. Then a causal G\"{o}del-type solution is permitted in k-essence theory for an electromagnetic field as matter content, since $E_0^2<0$ allows $m^2=4\omega^2$. Thus a causal G\"{o}del-type class of solutions of these equations that satisfies the condition $E_0^2<0$ for any function $K(\phi,X)$ is given by
\bea
m^2&=&4\omega^2,\label{59}\\
K(\phi,X) &=& \frac{m^2}{2\beta^2}\\
2K_x(\phi,X) \dot{\phi^2} &=& \frac{m^2}{\beta^2} + E_0^2.
\eea
It is important to emphasize that, eq. (\ref{59}) implies $r_c\rightarrow \infty$ and as a consequence the CTCs are not allowed in this class of solution. Thus, the k-essence theory allows both causal and non-causal G\"{o}del-type solutions.

\section{Conclusions}

K-essence is a theory that explains the accelerated expansion of the universe. It is minimally coupled to the scalar field with non-canonical kinetic energy. In this paper the behavior of the G\"{o}del and G\"{o}del-type solutions within this gravitational model is verified,  looking for the consistency of these metrics within such theories, and or their corresponding physical interpretations.  The nontrivial property of these solutions consists in breaking of causality implying in the possibility of the CTCs in this space-time. Our results have shown that: (i) the G\"{o}del metric is a solution in k-essence theory. The usual conditions of general relativity are modified due to the k-essence function $K(\phi, X)$. (ii) Considering the perfect fluid as matter source a non-causal G\"{o}del-type solution is found. Then the critical radius (beyond which the causality is violated) has been calculated. It depends on both k-essence theory and the content of matter. (iii) By taking a combination of perfect fluid with an electromagnetic field, and simply a single electromagnetic field the causal regions are allowed. Therefore, the k-essence theory accommodates both non-causal and causal G\"{o}del-type solutions. In addition, our results are for any function $K(\phi, X)$.

\section*{Acknowledgments}

This work by A. F. S. is supported by CNPq projects 308611/2017-9 and 430194/2018-8; J. G. da Silva thanks CAPES for financial support.

\end{document}